# Direct Optical Visualization of Water Transport across Polymer Nano-films

Swathi Suran*, Manoj M. Varma*†

*Centre for Nanoscience and Engineering, Indian Institute of Science, Bangalore, India

†Robert Bosch Center for Cyber Physical Systems, Indian Institute of Science, Bangalore, India

Email: swathi.suran@cense.iisc.ernet.in, mvarma@cense.iisc.ernet.in

**Abstract**

Gaining a detailed understanding of water transport behavior through ultra-thin polymer membranes is increasingly becoming necessary due to the recent interest in exploring applications such as water desalination using nanoporous membranes. Current techniques only measure bulk water transport rates and do not offer direct visualization of water transport which can provide insights into the microscopic mechanisms affecting bulk behavior such as the role of defects. We describe the use of a technique, referred here as Bright-Field Nanoscopy (BFN) to directly image the transport of water across thin polymer films using a regular bright-field microscope. The technique exploits the strong thickness dependent color response of an optical stack consisting of a thin (~25 nm) germanium film deposited over a gold substrate. Using this technique, we were able to observe the strong influence of the terminal layer and ambient conditions on the bulk water transport rates in thin (~ 20 nm) layer-by-layer deposited multilayer films of weak polyelectrolytes (PEMs).

## Introduction

Desalination using ultra-thin membranes including 2D materials is being increasingly explored recently [1-5]. Layer-by-layer assembled ultra-thin polyelectrolyte multi-layers (PEMs) [6-8] represent a versatile class of thin film system with widely tunable properties [9]. Physiosorption due to electrostatic interaction between the polycations and polyanions forms the basis of this deposition technique. It is well known that the properties of these multi-layers are affected by extrinsic factors such as salt ion concentration, pH and temperature which allows thickness tunability and sensitivity to a wide range of factors [10- 17]. This dependency allows PEMs to be used as sensors, optical coatings, polymer capsules for targeted drug delivery and nano-filtration membranes to mention a few applications [18 - 22]. Recently, polymer based nano-membranes are being explored for water filtration and desalinization [23, 24] making it increasingly important to study the water transport at nano-scales. In early 2000s, Bruening's [25-27] and Tieke's [28, 29] groups studied the use of PEMs as nano-filtration membranes. Detailed understanding of hydration and water transport mechanisms through PEMs are necessary for efficient application of PEMs in water filtration and desalination technologies.

In the past, extensive efforts have been made to study PEMs and its internal characteristics to better understand the hydration mechanism under different environmental conditions [30 - 32]. Poly (allyamine hydrochloride), (PAH) / Poly (acrylic acid) (PAA), Poly (allyamine hydrochloride), (PAH)/ Poly (sodium 4-styrenesulfonate), (PSS) and Poly (diallyl dimethyl ammonium chloride), (PDDAC) / Poly (sodium 4-styrenesulfonate), (PSS) have been the most investigated polymer systems. Unlike strong PEs, weak PEs such as PAH and PAA multilayer properties strongly depend on pH of the assembly solution [33 - 39]. An interesting feature observed in many of the hydration and water transport studies is the odd-even effect, i.e. the variation of PEM properties such as wettability, on the terminating layer [17, 33]. PEMs with odd number of layers are fabricated by terminating the PEM with a polycation/positive polyelectrolyte while an even numbered PEM is fabricated when the terminating layer is a polyanion/negative polyelectrolyte. Different techniques have been used to measure the hydration/immobilization of water in these odd-even polymer films. Contact angle measurements were used to study the surface wettability and but did not give inputs into wettability in the bulk [33, 38]. Methylene Blue (MB) penetration studies were carried out to study the extent of interpenetration and the influence of the underlying PE layers [39]. NMR techniques were used to quantify the degree of swelling induced due to hydration [40 - 42]. Additionally, x-ray and neutron reflectivity techniques have also been used to study the internal arrangement and water uptake at the surface and bulk of PEMs [43, 44]. Ellipsometry measurements have been used to measure the optical changes before and after hydration when the swollen film will cause a change in the bulk refractive index. Often water content in the PEMs is calculated by thickness change measured before and after swelling using an Ellipsometer. It is reported that hydration/water uptake in PEM is a bulk characteristic but governed by the surface charge of the terminating layer [45 – 48]. The techniques used in the past, though providing significant insights into hydration of PEMs, have been generally complex and do not provide a direct measurement of water transport rates and associated dynamics. Specifically, these techniques cannot measure the odd-even effect and in general a relative difference directly on the same sample. Additionally, these techniques cannot probe the role of defects or more generally the micro and nano-scale organization of the thin film on the bulk transport behavior.

In this article, we describe a technique which provides direct microscopic visualization of water transport in PEMs (as well as other polymer thin films) using simple bright-field optical microscopy. The measurement technique described here is based on the strong thickness dependent color contrast of a water-soluble germanium (Ge) thin film (~25 nm thick) deposited on optically thick gold [49]. Differences in water transport rates, for instance between a cationic terminated PEM and an anionic terminated PEM, leads to a difference in color between the two regions. The rate of color change can be converted to a water transport rate transverse to the film thickness (when lateral etching of Ge is negligible). Using this technique, we observed significant differences in water transport rates in polyelectrolyte multilayer films depending on the surface termination of the PEM for 3 different PEM systems including weak and strong PEs. The difference in water transport rates could not be explained by difference in hydrophobicity. Liquid immersion Atomic Force Microscopy (AFM) eliminated the possibility of differential swelling upon hydration leading us to conclude that the differential water transport between odd and even terminated PEMs is the result of differential hydration of the top-most PE layers. Essentially, the top layer presents a barrier to the entry of water molecules into the bulk. In the case of PAA-PAH system we showed that this barrier is a function of ionic strength and the odd-even effect can even be suppressed with an appropriate choice of ionic strength. The technique described here is generally applicable to measure relative transport difference or to probe the relationship between film structure and bulk transport and thus is expected to be of wide utility for transport studies through ultra-thin polymer membranes including single layer 2D sheets [49].

**Materials and Methods**

Weak polyelectrolytes PAH, Poly (allyamine hydrochloride) and PAA, Poly (acrylic acid) with an average Mol. Wt. of 17,500 and 450,000 and strong PEs PDDAC, Poly (diallyl dimethyl ammonium chloride) and PSS, Poly (sodium 4-styrenesulfonate) with an average Mol. Wt. of 200,000- 350,000 and 70,000 respectively, were used in this study. Both strong and weak PEs were procured from Sigma Aldrich. All the polyelectrolyte assembly/dipping solutions were made in 0.1 M NaCl solutions in 18MΩ cm Millipore deionized water, maintained at pH 5.5. In this pH regime, the PAH/PAA system has an exponential growth whereas the PSS/PDDAC system has a linear growth mechanism. Optical images were captured in upright Bright-Field microscope, Olympus BX 51M and color camera model DP73. The thicknesses were measured using Atomic Force Microscope by Bruker, the Dimension Icon tool. For polymers in the dry state, the thicknesses were measured using the tip, TESPA-V2 (Bruker) in the tapping mode and in the wet state, Scan Asyst Fluid tips (Bruker) were used in the Fluid mode scan.

Polyelectrolyte multilayers were prepared by layer-by-layer deposition method established by the group of Decher [6, 8]. Polyelectrolytes were coated on to the substrate by manually dipping the substrate in polycationic and polyanionic dipping/assembly solution for 1 minute alternately, with intermittent rinsing in DI water as shown in Fig.1 (a). The substrate (over which the PEM was coated) was a thin film structure consisting of a germanium film with thickness of about 25 nm deposited over optically thick (> 100 nm) gold films on a Silicon wafer. All depositions were carried out in a Tecport Sputtering tool.

Polyelectrolytes with both even and odd termination were fabricated by the LbL process. Measurement of water transport through these polymer films was made by imaging through a water immersion objective (60x, NA = 1) in a regular upright optical microscope. In order to measure relative difference (odd-even difference) in water transport through odd and even terminated polymer membranes simultaneously, a polycation/polyanion was drop casted on even/odd terminated PEM as shown in Fig. 1(b). A careful rinse of this drop casted area with DI water, provided us a sharp interface between an odd and even termination of the same bilayer. Both odd and even terminated sides of the PEM were observed in the same field of view using the water immersion lens with the sample immersed in a 0.02% v/v of $H_2O_2$ (Hydrogen Peroxide) diluted in DI water. The addition of hydrogen peroxide increased the etch rate of germanium. We refer to the $H_2O_2$: DI water solution, simply as water in this article.

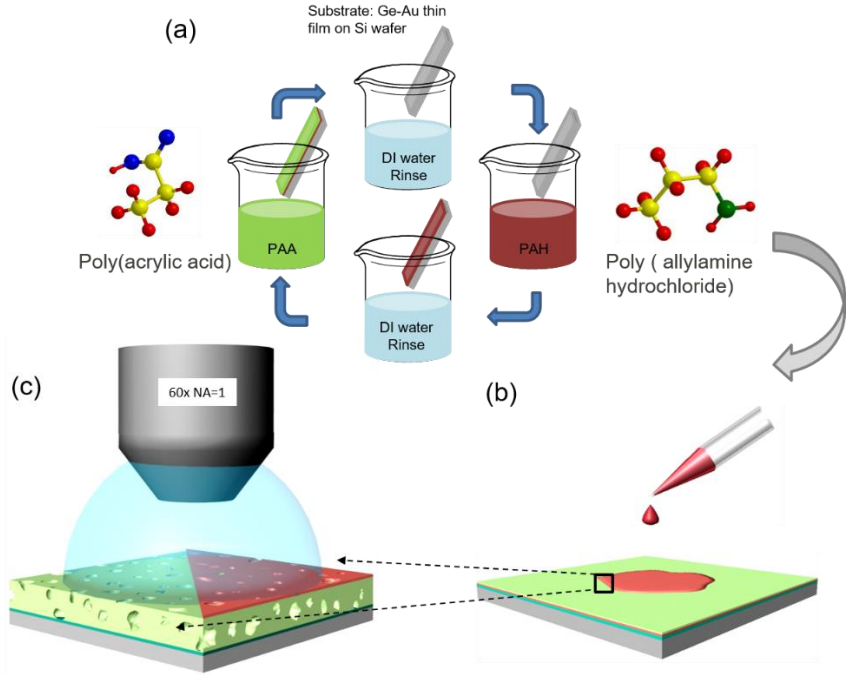

Figure 1. (a) Self-assembly of PEs due to electrostatic interactions. Layer-by-Layer deposition of positive and negative PEs with DI water rinse after each dip in the respective PE solution. This cycle can be repeated any number of times to get the desired number of bilayers. (b) Drop casting a positive PE on a negatively terminated PEM to create an interface for comparison of transport. (c) Observing the PE interface under the water immersion lens.

**Imaging water transport using Bright Field Nanoscopy (BFN)**

For (PAH/PAA)$_{6/6.5}$ system, it was observed that the polymer membranes terminated with positive polyelectrolytes (PAH terminated) facilitated a faster etch of the underlying Ge thin film when compared to PAA termination. Figure 2(a) displays the optical measurement of transport of water through PEs seen using a bright-field microscope as described earlier. The interface of with two surface terminations was observed in one field of view. It was observed that as soon as the water droplet is dispensed on top of the PEM, the side terminated with the positive polymer provided a faster and unobstructed transport of water causing the Ge beneath this layer to etch faster and reach till the Gold layer where the etching process terminated.  Hydrogen Peroxide which is added to DI water forms a eutectic/homogeneous mixture with $H_2O$ [Refer SI, Section-

1]. The etching process of Ge in the presence of water and hydrogen peroxide can be represented as [50, 51]:

$$\begin{rcases} Ge + 2H_2O_2 \rightarrow GeO_2 + 2H_2O \\ GeO_2 + H_2O \rightarrow H_2GeO_3 \\ GeO_2 + OH^- \rightarrow HGeO_3^- \end{rcases} \quad (1)$$

The relative difference in the etch rate of odd and even terminated PEMs are depicted in Fig. 2. All optical images depicting the water transport through polymers were captured at an interval of 10 seconds. Each image displayed in Fig. 2 are separated by 100 seconds in time. Red and green color channel values were extracted from each image in the series and plotted as a function of time (Fig.2 (b)).

A significant lag in the respective channels is observed for the PEM with PAA termination when compared to PAH terminated PEM. The color distance, D (t) is calculated using a simple Cartesian equation given by

$$D(t) = \sqrt{(R_t - R_f)^2 + (G_t - G_f)^2}$$

Here, $R_f$ and $G_f$ are the final values in the red and green channels whereas $R_t$ and $G_t$ are the values at any time 't' in the red and green channels respectively. The difference in the blue channel was negligible and hence has been neglected here. The color distance plot in Fig. 2(c) clearly shows the difference between the oppositely terminated surfaces. PAH terminated PEM region reaches the gold layer faster indicated by a constant value, as gold layer acts as the etch stop. It was also observed that all the PAH-terminated PEMs, irrespective of the thickness, etched Ge at the similar rates. Surprisingly, this rate was close to the etch rates of a bare Ge substrate [Refer SI, Section 2].

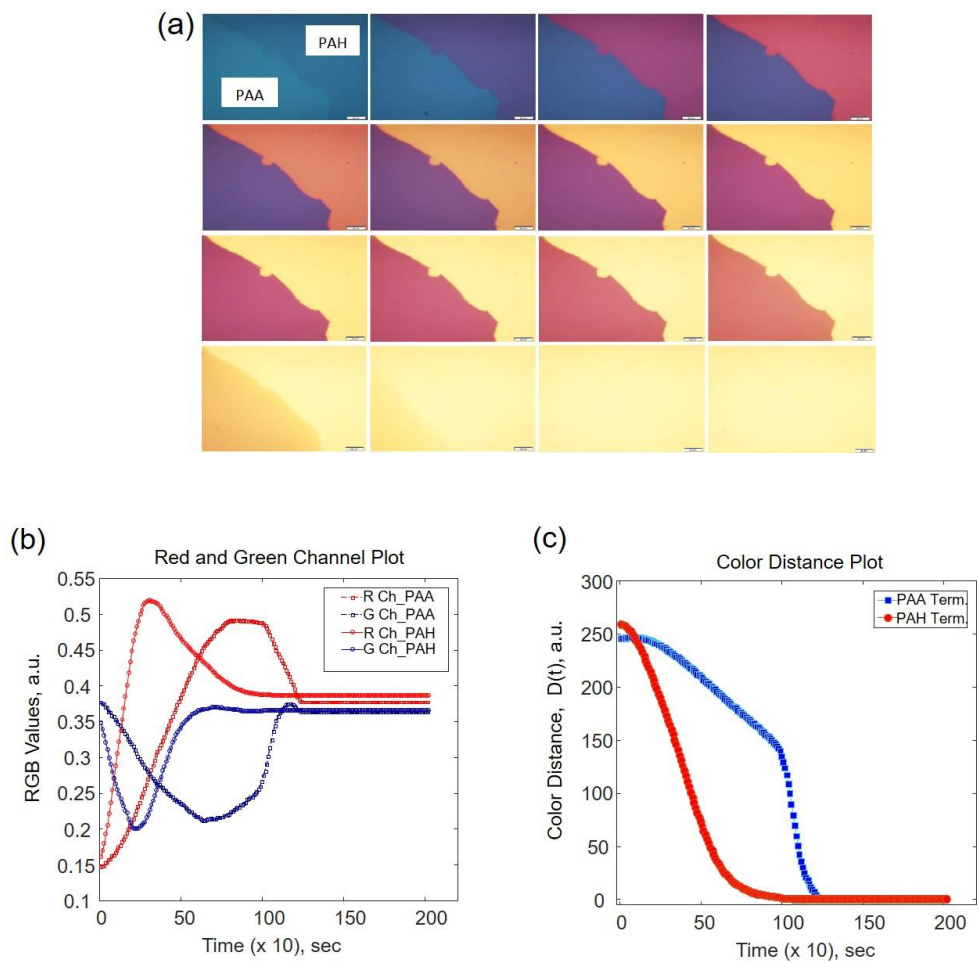

Figure 2. Imaging water transport in polymer membranes. (a) Time lapse bright-field images of the interface of a (PAH/PAA)$_{6/6.5}$. The region with positive and negative termination is marked with PAH and PAA respectively in the first image in the series. (b) The red and green channels values from both PAH and PAA terminated regions of interest are plotted as a function of time. (c) Color distance plot for PAH and PAA terminated PEM shows that the Ge under the PAH region erodes faster reaching the Gold film

We also studied relative water transport difference in other polymer systems, namely, PSS/PAH were both PEs were equally charged at pH 5.5 and PSS/PDDAC which are not affected by the pH of the dipping solution. We define a term, $\Delta$ Etch Time, as the difference between the time it takes to etch Ge from beneath the negatively terminated and positively terminated PEMs. Δ Etch Time Vs average etch durations for different polymer systems have been plotted in Fig. 3. It was observed that the strong PE system took approximately 10 times longer to etch Ge owing to its dense conformations of the layers [52, Refer SI, Section 3]. The change in color of Ge as it etched can be directly translated to find the volumetric flux of water through these PEMs. From the empirical data, the volumetric flux was calculated to be about 150 pL/min.cm$^2$ and 240 pL/min.cm$^2$ for a PAA and PAH terminated (PAH/PAA)$_6$ bilayer system Refer SI text, Section 4.

A remarkable feature noticed was that the Δ Etch times are independent of the number of bilayers for the PAH/PAA system, which rules out the possibility of swelling effects in polymers in these thickness regimes.

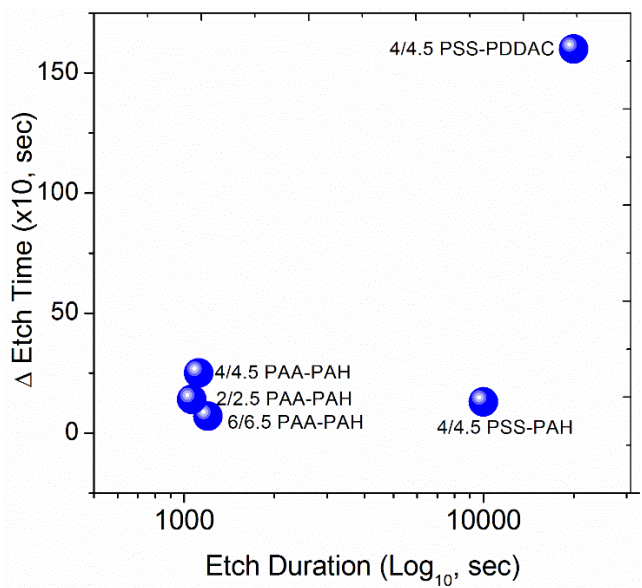

Figure 3. Δ Etch time plotted against the average etch durations for different polymer systems.

**Surface wettability and hydration dynamics studies**

Surface wettability is a characteristic of the surface forces available at the interface. Sequentially adsorbed polyelectrolytes alter the net wettability of the PEM. It was reported previously that the surface charges associated with the PEs affects the wetting kinetics and hence alter the contact angle of water [53]. In this study, contact angle (CA) measurements carried out to study the surface wettability revealed that PAH capped PEM was relatively less hydrophilic compared to a PAA capped PEM, with a contact angle difference of about 6-8° on an average (Fig. 4). This rules out the possibility of the odd-even effect arising out of hydrophobicity difference as PAA surface with slightly higher hydrophilicity should have transported water faster or at least comparable to that of PAH. We also studied the wetting dynamics and saw the contact angle of the sessile drop recede with time for both PAA and PAH terminated PEMs. A CA drop of 6° was observed for a PAH termination whereas a 3° drop was observed for a PAA terminated PEM surface when measured over a period of approximately 2 mins.

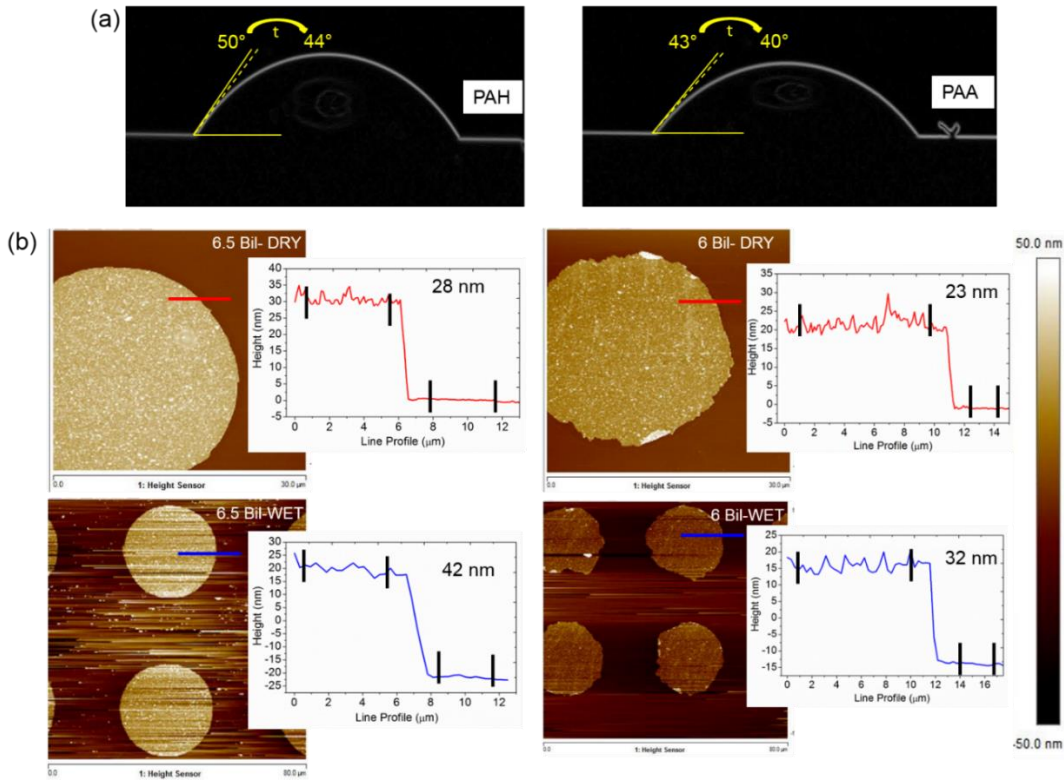

Figure 4. Surface and bulk wettability studies (a) Contact angle measurements of a sessile drop on a PAH and PAA terminated PEM system. Transient effects on the sessile droplet was observed over a period of t, 2 mins. The measured CAs are indicated in the respective figures. (b) AFM measurements in air and in fluid mode measured the swelling of the PEMs in presence of bulk water. The measured thickness values are provided in the inset of each scan.

In-order to investigate if the relative difference in water transport rate was due to a relative difference in swelling of the two polymers, Fluid mode Atomic Force Microscopy (Fluid-AFM) was performed to measure the swelling behavior of the PEs immersed in water. In the dry state, a 6 and 6.5 bilayer PAH/PAA measured about 23 nm and 28 nm. Subsequently, in situ swelling measurements were carried out by immersing the sample in water in the liquid mode AFM (for the same durations as in the optical experiments). In the wet state, post swelling thicknesses measured about 32 nm for PAA and 42 nm for PAH terminations respectively. For thinner PEMs, namely, (PAH/PAA)$_{2/2.5}$ and (PAH/PAA)$_{4/4.5}$ bilayers, there was no significant swelling while immersed in water for the same durations [Refer SI, Section 5]. This data showed that the increased physical thickness of PAH terminated layer does not determine the water transport rate as the PAH terminated layers etched faster than PAA terminated ones.

**Discussion and Conclusion**

Swelling experiments provide very useful insights into hydration of the bulk or in other words, internal wettability. Post saturation in water, the PAH terminated PEM swells considerably more than PAA terminated PEMs. It was observed that despite the relatively large initial thickness of the PAH terminated PEMs, the water transport was faster than a PAA surface terminated PEM. This confirmed that the physical thickness of the PEM layer did not have significant influence on the transport of water molecules. Moreover, from the dynamic CA measurements, it was inferred that the PAH terminated PEM allowed a faster and more efficient water uptake. Therefore, differential swelling can be ruled out as a mechanism for the differential water transport (etch) rates of Ge.

This observation led us to look for a mechanism based on difference in surface impedance or permeability to water transport between the PAH and PAA terminated layers. It has been reported that, PAH has a charge density ($\xi$) greater than 1, which is due to closely spaced electrolyte units on the polymer chain [54]. Hence, unlike the bulk of the PEM where all the positive and negative binding sites of the PEs are intrinsically paired, the surface provides a significant number of uncompensated binding sites when the surface is terminated with a PAH layer. These positive binding sites are satisfied by the counterions made available from the dipping/ bathing solution. The $Cl^-$ ions from the dipping solution containing NaCl extrinsically binds to these PAH binding sites leading to a gradient of $Cl^-$ ion concentration and the consequent osmotic pressure drives water into the bulk PEM due to osmosis. The presence of $Cl^-$ ions on the surface makes the PAH terminated PEM more permeable to water compared to PAA terminated PEMs [More info, Refer SI text, Section 6]. Additionally, when the surface is terminated with PAA, the uncompensated carboxylic acid groups available on the surface form hydrogen bonds with water molecules due to their strong affinity [55] whereas the amine groups on a PAH surface do not form $H_2O$ clusters. In order to test our hypothesis regarding the surface barrier model described above, we conducted the etching experiments with water containing different concentrations of salt (NaCl) serving as a reservoir deposited over the PEM. As mentioned previously, the PEM layers were deposited at 0.1M NaCl. Etching experiments were conducted with 0.01M, 0.1M, 0.5M and 1M NaCl. $H_2O_2$ was not added into the etching solution to avoid possible reactions. With the high ionic strength solutions (0.5M and 1M), we did not observe any visible color change even after long etching times. This is because, the external ionic strength is much larger than the ionic strength used for LbL assembly. In this case osmotic pressure would prevent the uptake of water. Due to the extremely small etch rates no measurable odd-even effect was observed at 0.5M and 1M etching conditions. At 0.1M the reservoir and the bulk PEMs are at similar ionic strengths and even in this case the odd-even effect is effectively suppressed. It is only when the reservoir ionic strength is below that of the ionic strength used in LbL assembly that we start seeing the odd-even effect. Note that the data in Fig. 2 showing strong odd-even effect is obtained at 0M (no salt). The odd-even effect as a function of the ionic strength of the reservoir is summarized in Fig. 5 where the difference in etch times between PAH and PAA terminated PEMs are plotted against ionic strength. [More info, SI text, Section 7]. A mathematical model based on the generic framework of diffusion across a surface barrier (which may arise from ionic gradients, surface molecular bonding and so on) is capable of producing odd-even effects observed in PAA/PAH and other polyelectrolyte systems [SI text, section 8].

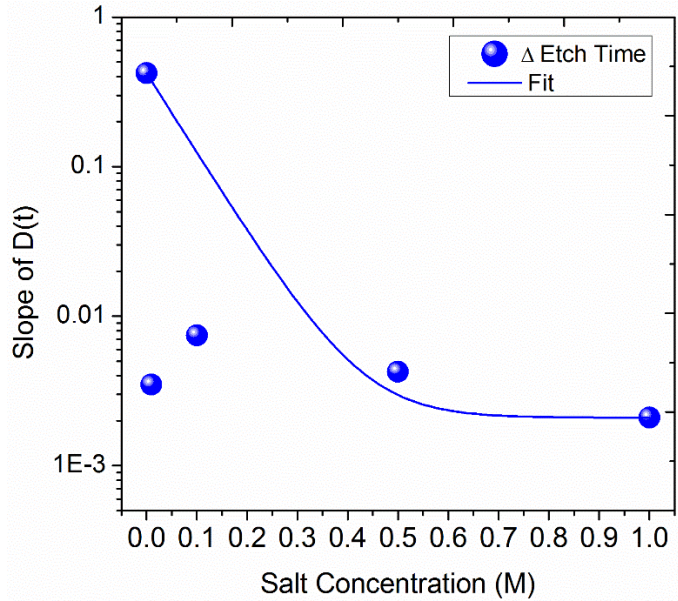

Figure 5: Difference in etch times between PAH and PAA terminated PEM surfaces is plotted for different salt concentrations of the etching solution.

In the case of PAA/PAH we explored this surface barrier term in details and showed its dependence on ionic strength. Similar mechanisms may be suggested for other systems.

To summarize, we presented a technique for direct visualization of water transport across nano-membranes using bright field microscopy. Using this technique, we were able to probe the difference in water transport rates across PEMs depending on the terminating layer. Focusing on the weak polyelectrolyte system PAA/PAH, we showed the significant role of ionic gradients in the creation of the observed odd-even effect. Finally, we showed that a mathematical model involving a surface permeability term is sufficient to capture the observed odd-even effect. We believe that the technique described here will be of wide utility for water transport studies through nano-membranes, particularly in studying the effects of film structure on bulk transport in detail.

# SUPPLEMENTARY INFORMATION

*Section 1: Hydrogen Peroxide with water.*

Hydrogen peroxide is an oxidant and a prime etchant of Germanium. Etching of Ge occurs by oxidation of Ge by $H_2O_2$ followed by dissolution of the oxidation products in aqueous solution. In our experiments, we use an aqueous solution containing 0.02% v/v of $H_2O_2$. When dissolved in $H_2O$, $H_2O_2$ forms a eutectic mixture due to hydrogen bonding between them. If not in the presence of catalysts, prolonged exposure to factors like temperature and pH can self-decompose $H_2O_2$ to water and oxygen.

*Section 2: Etch rate of Bare Ge.*

The color distance map of transport of water through both negative and positive termination for a PAH/PAA system for different thicknesses is plotted in Fig. S1. It was observed that the underlying Ge etched in similar time scales for all the positively terminated PE surfaces irrespective of the thickness. Their profile and duration matched closely with that of a bare uncoated Ge sample.

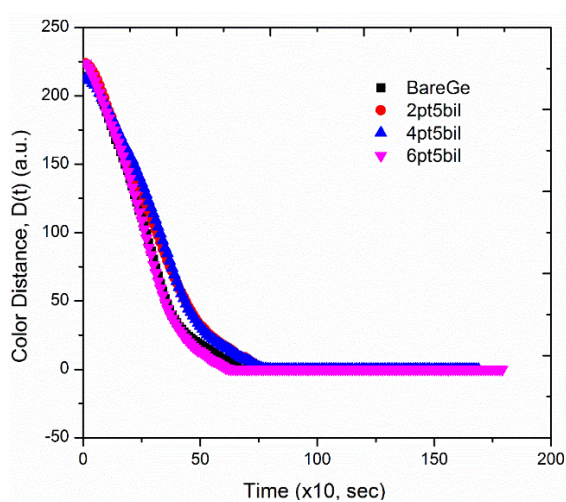

Fig. S1. Color distance plot for (PAH/PAA) system for PAH/odd terminated bilayers.

*Section 3: Stronger PEs have longer etch durations*

Experiments with strong PEs clearly demonstrated that it took longer times to etch Ge completely which in turn meant that the transport is slower in these systems. The odd-even surface effect did persist in these systems irrespective of the denser PE conformations. However, the experiment time scales are an order of magnitude higher than the experiment time scale for the weak PEs. A color distance plot of a (PDDAC/PSS)$_{4/4.5}$ bilayer illustrates the large difference in time scales.

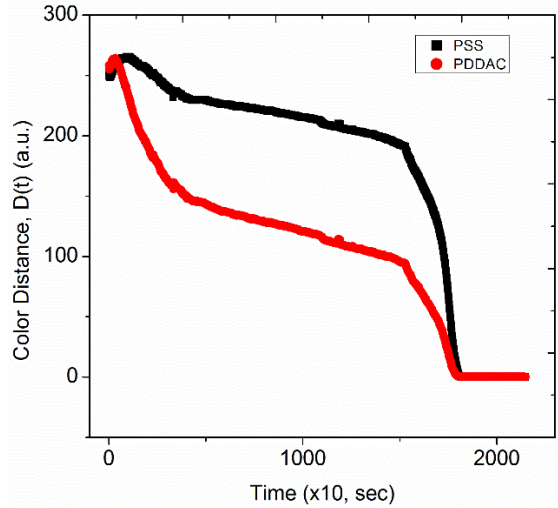

Fig S2. Color distance plot for a (PDDAC/PSS)4/4.5 bilayer.

*Section 4: Calculating Volumetric flux*

From the experimental observations in Fig. 2 in the main text, the change in color which originated due to the change in the thickness of Ge is a direct translation of the volumetric flux of etchant/ water reaching the thin film. Assuming that for every one H$_2$O molecule, one Ge atom is removed, the volumetric flux can be given by this formula,

$$V_{H_2O} = \left(\frac{h_{Ge} \times \rho_{Ge}}{m_{Ge}}\right)\left(\frac{m_{H_2O}}{\rho_{H_2O}}\right)$$

The $\rho_{Ge}$ and $\rho_{H_2O}$ are the densities of the Ge and H$_2$O, $m_{Ge}$ and $m_{H_2O}$ are the masses of Ge and H2O respectively. $h_{Ge}$ is the etch rate of Ge for the individual terminated PEMs. Based on the reactions determining dissolution of Ge, the water transport difference due to the presence of a single layer of PAA can be calculated to be around 0.88 µL/min.m$^2$. In relative terms, PAA impedes the water transport by about 60%.

*Section 5: Swelling experiments on thinner bilayers of PAH/PAA.*

Thickness of (PAH/PAA)$_{2/2.5}$ and (PAH/PAA)$_{4/4.5}$ bilayer, both in the dry and wet state were measured using the Atomic Force Microscope in the regular soft Tapping mode and Scan Asyst fluid mode. In thinner bilayers, the thicknesses of the PEMs fall under the influence of the substrate as reported by Tanchak and Barrett [S1]. Our results comply with these reports as seen in Fig. S3 and S4. The 2x2 panel shows the thicknesses of the PEM while in the dry state and when immersed in water. It is clear from the measurements that there is not significant increase in thicknesses of PEMs for both (PAH/PAA)$_{2/2.5}$ and (PAH/PAA)$_{4/4.5}$. Hence, it was clear that the swelling effect was observed only in the thicker PEM films and negligible in the thinner films.

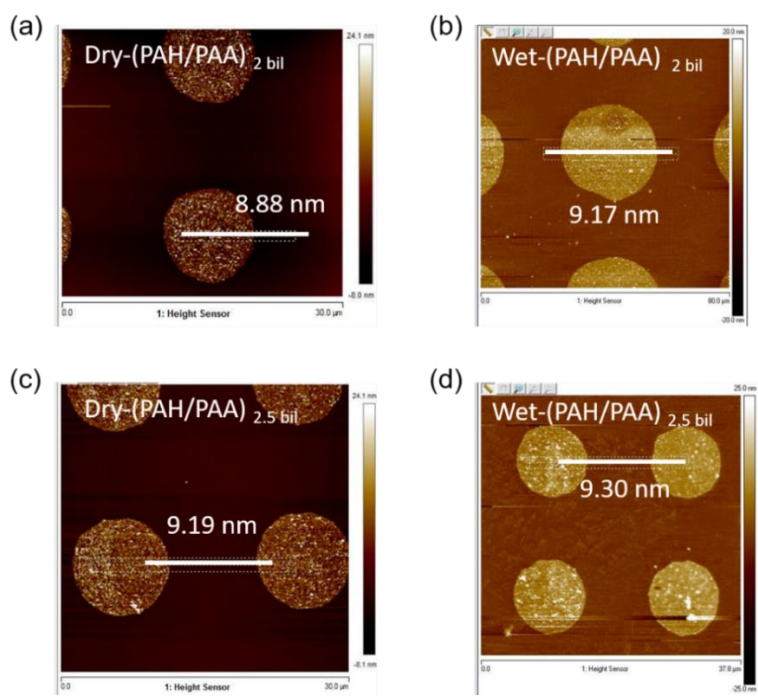

Fig S3. Dry and wet AFM images of (PAH/PAA)$_{2/2.5}$

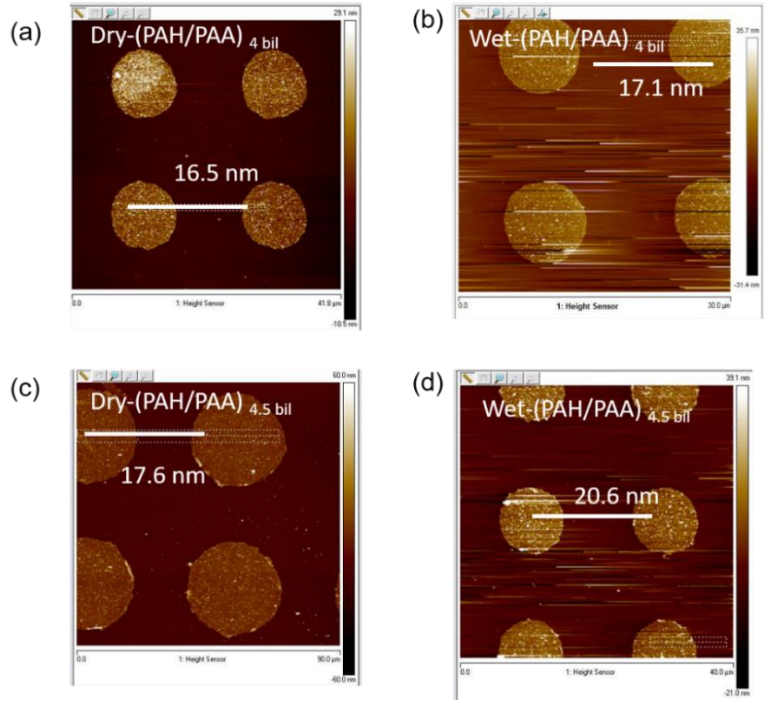

Fig S4. Dry and wet AFM images of (PAH/PAA) $_{4/4.5}$

*Section 6: Extrinsic Charges and Water Uptake*

PAH has a higher charge density than PAA, i.e. the distance between successive repeat units are smaller than Bjerrum Length [S2, S3]. Bjerrum Length is given by "Separation at which the electrostatic interaction between two elementary charges is comparable in magnitude to the thermal scale, $K_BT$ where $K_B$ is the Boltzmann constant and T is the absolute temperature. Manning's development of the counterion Condensation Theory treats the distribution of counterions around a highly-charged PE in terms of linear charge density parameter, $\xi$". Here, $\xi = \frac{l_B}{b}$ where $l_B$ is the Bjerrum Length given by, $l_B = \frac{e^2}{4\pi\varepsilon_0\varepsilon k_BT}$ typically measures about 7.31 Å @ 25°C in water. $'b'$ is the axial distance between the successive charge fixed on the monomers in the PE chains. For PAH, the $b$ is very small, smaller than $l_B$ and $\xi$ turns out to be >1. Most of all other PEs used have a larger $b$ than $l_B$ and hence have a charge density, $\xi$ <1. NaCl in the dipping/bathing solution dissociates to Na+ and Cl- which compensates binding sites on PEs if not paired by the opp. PE during the multilayer formation. Na+ and Cl- are called surface counterions. The Fig. S5 shows the higher charge density on PAH which are compensated by the Cl- . Polymer charges within the bulk of the multilayer balance with 1:1 stoichiometry termed "Intrinsic" charge compensation.

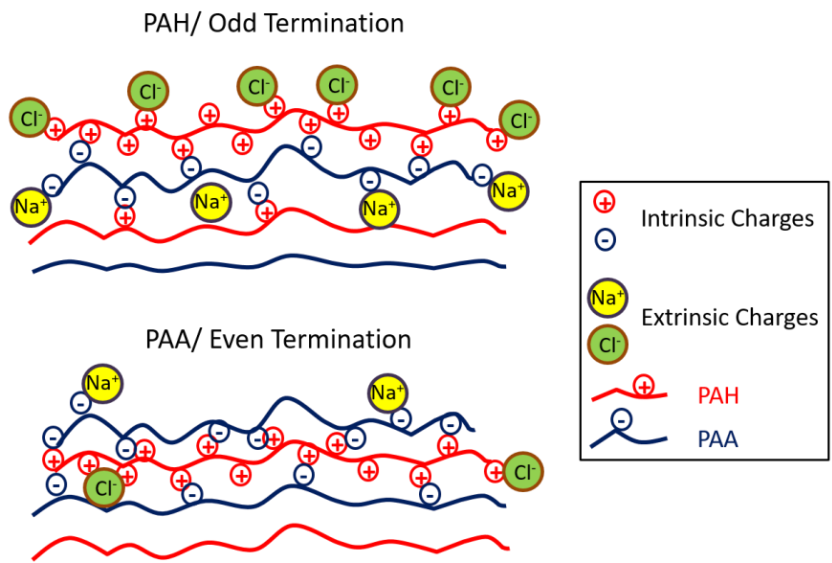

Fig. S5: Extrinsic and Intrinsic charge compensation in a PAH/odd and PAA/even terminated (PAH/PAA) system.

At a PAH terminated surface, the charges are overcompensated/unbalanced/not neutralized. Hence the PE charge stoichiometry is 'not' 1:1. Uncompensated binding sites of PAH surface are then neutralized by Cl- counterions- "Extrinsic" charge compensation. In Bulk water, larger presence of Cl- ions at the surface (of a PAH terminated PEM) induces an osmotic pressure (OP). Increasing OP, causes osmosis of water from the surrounding liquid in the PEM. PAH surface acts as permeable membrane with large values of $\rho$.

*Section 7: Response of a PAH terminated surface to different ionic concentration*

The unbalanced PAH surface termination compensated by the Cl- ions had an ionic concentration of 0.1 M which is the concentration of the PE in the solution form. It is by osmosis (higher ionic concentration at the PAH surface than the surrounding solution) that water transports into the PEM bulk. By varying the ionic concentration of the etching or immersion solution, the transport through the PEM can be altered. To study this, a PAH / PAA interface was studied for different concentrations of NaCl in the etching solution ($H_2O$). Due to possible exothermic reactions, $H_2O_2$ was avoided in the etching solution. Four different concentrations, 0.01 M, 0.1 M, 0.5 M and 1 M NaCl (in $H_2O$) solutions where prepared to be used as the etching/immersion solution. With a 0.01 M NaCl concentration of the etching solution, the transport was found to be faster when compared to 0.1 M, 0.5 M or 1 M solutions. Having a smaller ionic concentration (0.01 M) relative to the PAH surface (0.1 M), the rate of water transport into the PEM bulk was larger which was as expected. However, at equivalent concentrations with both PAH surface and immersion solution at 0.1 M the transport rates were found to be 35% slower than when the solution was maintained at 0.01 M. Ideally, there shouldn't have been any movement of water if the concentrations were similar. A slow transport even in this condition could mean that the ionic concentration is not even throughout the PAH surface. At higher concentration, 0.5 and 1 M, the transport was more or less seized as seen in Fig. S6.

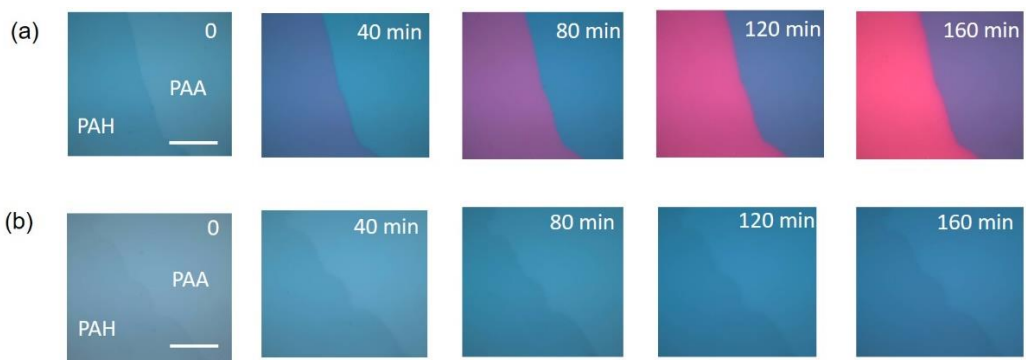

Figure S6: The transport difference across PAH and PAA terminated PEM when the immersion solution had salt concentration of (a) 0.01M, (b) 0.5M

*Section 8: Diffusion equation with a permeable wall*

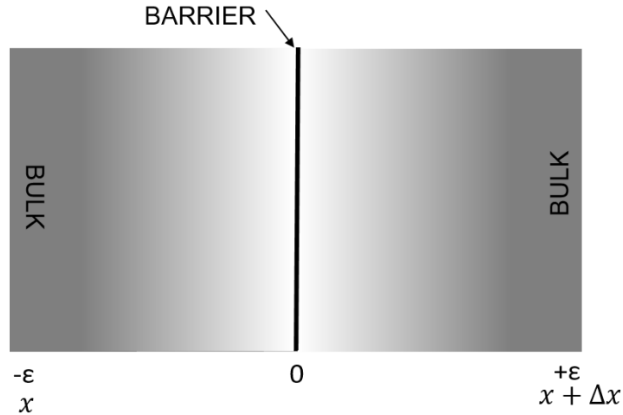

Figure S7: Diffusion model with a barrier wall

We mathematically modeled the transport of water through the PEM as a diffusion problem with a surface barrier as shown in Fig. S7 [S4-S6]. This problem has been analyzed by Tanner in great detail [S5]. The two key parameters in the model are the diffusion coefficient, D and the permeability coefficient, rho ($\rho$). Diffusion coefficient, D associates with the bulk of the PEM whereas the permeability coefficient, rho ($\rho$) relates to the surface of the PEM. By presenting two different values for the permeability, different surface terminating layers will lead to different water transport rates even with the same bulk diffusion values. In the case of the PAA/PAH system we identified difference in permeability with the Cl- ion gradient formed with PAH termination and the increased molecular bonding of water with PAA. The cumulative effect of both of these is to increase the permeability value for PAH termination relative to PAA. Following Powels [S6], we derive the concentration profile due to diffusion with barrier as

$$C(x,t) = \rho C_0 \int_0^\infty e^{-2\rho x'} \, erfc\left(\frac{x + x'}{2\sqrt{Dt}}\right) dx'$$

We assumed that the etch rate of Ge would be proportional to C (x,t) computed above from which we can compute the height of Ge film as a function of time. As the initial thickness of Ge used in these experiments was 25 nm, we can use this height profile to estimate the etch time (defined as the time at which Ge film thickness reaches zero). We numerically computed the etch times for different values of D and rho ($\rho$). Diffusion coefficients used in the simulation were taken from previously published data [S7-S9]. We then plotted the ratio of the numerically computed etch time and the experimentally observed etch time for both PAH and PAA terminations as shown in Fig. S8 (a) and (b).

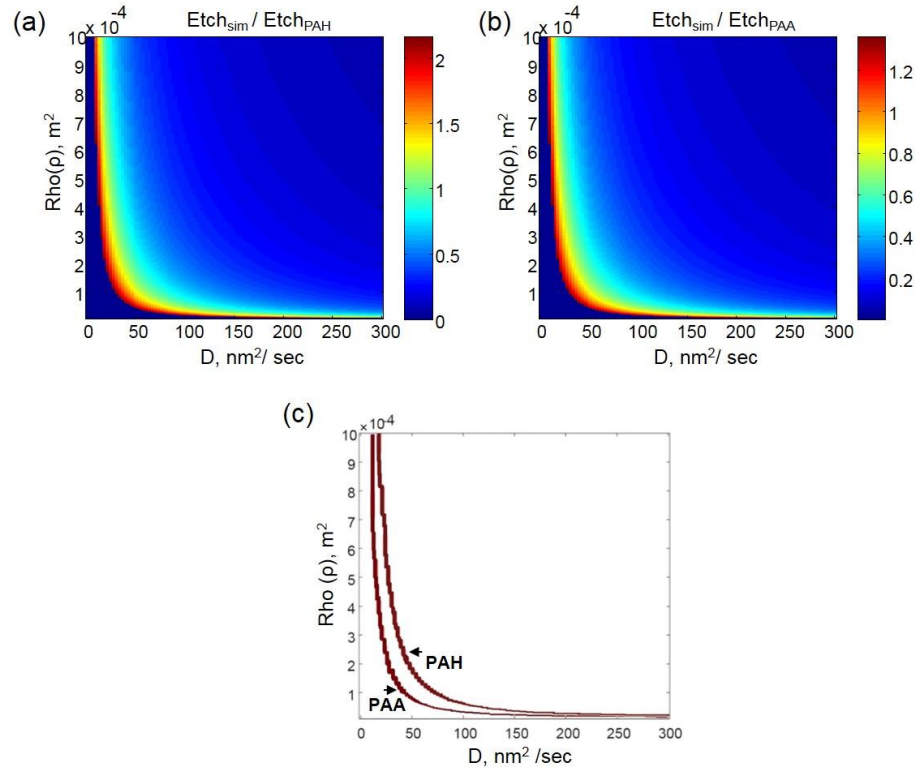

Figure S8: Ratio of the simulated and experimental Ge etch times for PAH/odd and PAA/even termination PEMs are plotted in (a) and (b). (c) Rho Vs D plot for $(PAH/PAA)_{6/6.5}$ system. This plot is obtained by thresholding values of plot (a) and (b) to either 1 or 0 (with 2% tolerance) which signifies complete etch of Ge and no etch of Ge respectively.

Figure S8 (c) shows the choice of D and ρ which is consistent with experimental observation. The curves corresponding to these parameters for PAA and PAH are obtained from the locus of points in Fig. S8 (a) and (b) whose value is equal to one. For a given value of bulk diffusion coefficient a slightly lower value of permeability of PAA relative to PAH is sufficient to cause the observed lowering of etch rate.

**SI References**